\documentclass[multphys,vecphys]{svmult}

\usepackage{makeidx}         
\usepackage{graphicx}        
\usepackage{multicol}        
\usepackage[bottom]{footmisc}

\makeindex             

\begin{document}
\title*{Muon g-2 Constraints to SUSY Dark Matter over the Next Decade}
\author{P.~Cushman}

\institute{Physics Department, University of Minnesota, 116 Church St. SE\\
Minneapolis, MN 55455, USA
\texttt{prisca@physics.umn.edu}}
\maketitle

\begin{abstract}
The anomalous magnetic moment of the muon has been measured to 0.5 ppm in a series of precision experiments at the Brookhaven Alternating Gradient Synchrotron. \cite{g201,g202,g204} The individual results for each sign:\\ a$_{\mu}^+$ = 11 659 204(7)(5) $\times$10$^{-10}$ ~~and~~ a$_{\mu}^-$ = 11 659 214(8)(3) $\times$ 10$^{-10}$\\ are consistent with each other, so that we can write the average  anomaly as a$_{\mu}$(exp) = 11 659 208(6) $\times$ 10 $^{-10}$  (0.5 ppm).  A discrepancy, $\Delta$a$_{\mu}$, between  the measured value a$_{\mu}$(exp) and the Standard Model a$_{\mu}$(SM) is a signal for new physics.  Assuming that such a discrepancy is due to contributions from supersymmetric particles provides a framework which can be used to constrain the mass of the dark matter particles, assumed to be the lightest neutral supersymmetric particles.  The deviation from the standard model has varied between 1.5$\sigma$ and 3$\sigma$ significance, dominated by uncertainties in the hadronic contribution to the standard model calculation.  Currently the standard model prediction is calculated to 0.6 ppm precision and $\Delta$a$_{\mu}$ = 23.5 (9.0) $\times$10$^{-10}$ representing a 2.6$\sigma$ deviation.  We expect that the error on a$_{\mu}$(SM) will be reduced by a factor of two within the next decade.  To fully utilize this improvement, a new g-2 run is proposed for the near future.  If the mean $\Delta$a$_{\mu}$ remains the same, this would result in close to a 6$\sigma$ discrepancy.  In this case, we would expect to see SUSY particles at the LHC and use the g-2 results to measure tan $\beta$.  If, instead, the Standard Model is confirmed to this precision, gauginos must have masses higher than $\sim$500 GeV/c$^2$ and simple SUSY dark matter models will be severely constrained. 
\end{abstract}

\section{ Historical Summary} 

Precision measurements are a complementary approach to investigating the highest energy, smallest scale frontier of particles and interactions. Over the last decade, E821, the Brookhaven g-2 experiment, has successfully mounted a precision challenge to the standard model. The magnetic moment is defined as $\vec{\mu} = g\frac{e\hbar}{2mc}\vec{s}$, where g is the gyromagnetic ratio. Deviations from a purely pointlike g=2 Dirac particle are characterized by the anomaly a=(g-2)/2. The anomaly for leptons is $\sim$10$^{-3}$ due to interactions with virtual particles which couple to the electromagnetic field, thus providing a laboratory for testing the Standard Model.  Whereas the electron anomaly provides the most precise measurement of the fine structure constant $\alpha$, the muon anomaly is more sensitive by m$_{\mu}^2$/m$_X^2$ to virtual W and Z gauge bosons, as well as any other, as yet unobserved, particles in the hundreds of GeV mass range.  

 The 2.6 $\sigma$ discrepancy announced four years ago sparked debate on the theoretical calculation and encouraged further work on reducing the uncertainty in the 1$^{st}$-order hadronic contribution.  One of the more startling developments, approximately 6 months after the announcement of the first precision result (1.3 ppm), was the revelation by the Marseilles group \cite{KNyff02} that one of the contributions to the Standard Model theory, namely the hadronic light-by-light term, had been independently assigned the wrong sign by at least two separate groups.  Kinoshita \cite{HayK01} and Bijnens \cite{Bi02} studied their previous work and found that they both had used an incorrect sign convention in a matrix evaluation in a widely-used computer program. This moved the theoretical value by 17 $\times 10^{-10}$ (by more than its stated uncertainty) in the direction of the E821 result, thus reducing the discrepancy to 1.6$\sigma$.  The next g-2 run, with an improved precision of  0.7 ppm, left the mean unchanged, but reduced the error bars, again indicating a 2.6 $\sigma$ discrepancy with the Standard Model.  

Meanwhile, in order to reduce the uncertainty on the hadronic correction, the use of vector spectral functions from the study of hadronic $\tau$-decays in ALEPH was introduced  by Alemany et al. \cite{Alem98}. Previously, the only handle on the hadronic vacuum polarization contribution at the low center of mass energies which are relevant for g-2 came from the dispersion relation: 

\begin{equation}
a_\mu^{had,1} \propto \int_{(2m_\pi)^2}^{\infty} ds~ \frac{K(s)}{s} R(s)
  ~~where~
~R(s) = \frac{\sigma(e^+ e^- \rightarrow hadrons)}{\sigma(e^+ e^- \rightarrow \mu^+
    \mu^-)}
\end{equation}                                                                                    

\begin{figure}[tbp]
\centering
\includegraphics[width=3.5in]{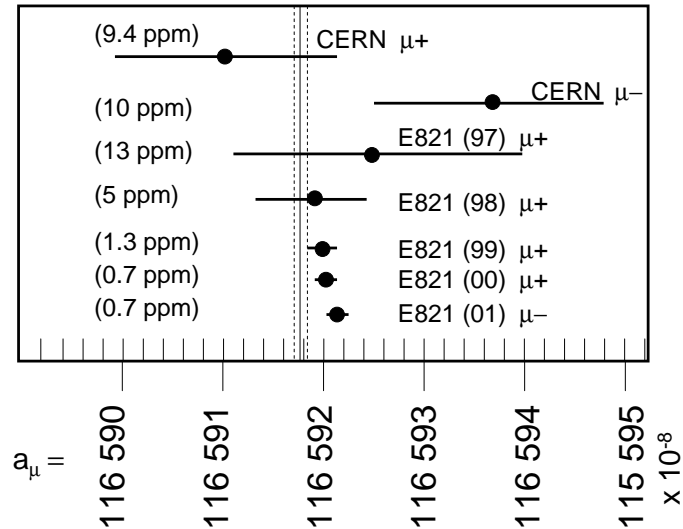}
\caption{An historical look at the sequence of g-2 results in improving precision.  The line represents the Standard Model calculation as of early 2004, using e+e- data to deduce the hadronic contribution.  The dotted lines indicate the uncertainty in the SM calculation.}
\label{exphistory}   
\end{figure}

R(s) is determined from a compilation of experimental results dominated by the CMD-2 experiment at Novosibirsk.  When the $\tau$-decay data were first combined with the e+e- data in 1998, it halved the error bars on the hadronic contribution to the SM calculation.  However, over the past 5 years the continued operation and analysis of CMD-2 has improved the e+e- data to such an extent that the two methods were found to be in disagreement with each other.  Thus, in order to quote a discrepancy with theory, it became necessary to distinguish which hadronic correction you were referring to.  In 2003 the Novosibirsk collaboration completely reanalyzed their $\pi\pi$ channel \cite{Akhmet03} following the discovery of a mistake in their normalization (the t-channel leptonic vacuum polarization contribution was missing in the Bhabha scattering cross section).  Their correction increased their published hadronic cross sections by 2.5\%, thus reducing, but not erasing the discrepancy between the two theoretical approaches (especially for energies above 0.85 GeV).  

New results from KLOE, BaBar and Belle can provide an independent method to distinguish between the two by using radiative decay to scan the center of mass energies in the region relevant to g-2 (so-called "radiative-return" method).  A precise measurement of the pion form factor has been reported by KLOE \cite{Alioso03}\cite{DiFalco03}.  It confirms the Novosibirsk e+e- result.  Preliminary results on  4-prong final states by BaBar \cite{Davier03} also bolsters confidence in the e+e- data. On the other hand, branching ratios from CLEO and OPAL continue to confirm the ALEPH data, thus indicating that the $\tau$-decay construction may be affected by a fundamental misunderstanding in how we apply CVC, isospin corrections, or the electroweak symmetry breaking. Ghozzi and Jegerlehner \cite{GhoJeg04} argue that by simply allowing the mass of the charged-$\rho$ to differ from $\rho^0$ is  sufficient to account for this. Davier \cite{Davier03} shows that even assuming this modification, a detailed comparison of the shape of the pion form factor reveals some discrepancy.  This in itself may be an indication of new physics.  Most are agreed, however, that in any comparison of a$_{\mu}$(exp) to a$_{\mu}$(theory), the direct result using e+e- data is more reliable at this time. 

The final g-2 experimental result was announced in January 2004 \cite{g204}. This was a 0.7 ppm result with opposite sign muons.  It was consistent with the previous data sets, despite reversing the magnetic field in the storage ring.  However, the mean value was slightly higher than the earlier value, serving more to emphasize than detract from any Standard Model discrepancy.  In figure \ref{exphistory}, the BNL g-2 results are shown together with those from the old CERN experiment.  The line represents the Standard Model calculation using e+e- data \cite{DEHZ03}.  Assuming CPT holds, combining all our experiments, and properly accounting for correlated systematics, the final experimental value for the anomalous magnetic moment is now at a$_{\mu}$(exp) = 11659208(6) $\times$ 10$^{-10}$  or a precision of 0.5 ppm.  This further precision also tends to increase the significance of the discrepancy, bringing us back to the original 2.6$\sigma$ significance. How$\Delta$a$_{\mu}$ may change over the next several years is now in the hands of the theorists until such time as a new g-2 experiment can be mounted.

\section{Current Experimental Status}
Pions produced on a nickel target were directed down a beamline which momentum selected the forward-going decay muons to produce a 96\% polarized muon beam. The muons were injected into the storage region via a superconducting inflector magnet. A pulsed magnetic kicker bumped the muons onto stored orbits in a uniform 1.45 T field and electrostatic quadrupoles provided vertical focusing.  The spin vector of the polarized muons precesses relative to the momentum vector with an anomalous frequency $\omega_a$, given by:

\begin{equation}
\omega_a \equiv \omega_S - \omega_C = a_\mu\,\frac{e}{m_\mu c}\,B ~~since~~\omega_S =\left[ 1 + \gamma\left(\frac{g-2}{2}\right)\right]
\underbrace{\frac{eB}{m_\mu c\gamma}}_{\omega_C}
\end{equation}

\begin{figure}[tbd]
\centering
\includegraphics[width=4in]{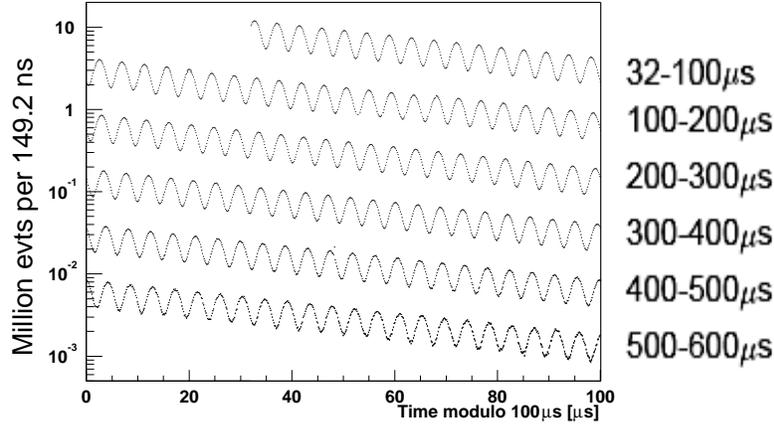}
\caption{The number of decay electrons detected by the calorimeters as a function of time after injection, showing the g-2 modulation superposed on the exponential decay of the parent muon (2001 negative muon data).}
\label{wiggle}   
\end{figure}

The dependence of $\omega_a$ on E was eliminated to first order by choosing a $\gamma$ which cancelled out the second term in equation 2, corresponding to a muon momentum of p = 3.094 GeV/c. The a$_{\mu}$ was then extracted from the ratio of the measured anomalous precession $\omega_a$ to the free proton precession frequency $\omega_p = \mu_p B/\hbar$ in the same magnetic field. The proton magnetic moment entered as the ratio $\lambda=\mu_{\mu}/\mu_p$ measured by the muonium hyperfine structure interval \cite{Liu99}. B was measured in situ every few days by a trolley with 17 NMR probes, and interpolated between trolley runs using $\sim$150 stationary probes.

To find $\omega_a$, the decay positrons (electrons) from $\mu^+ \rightarrow e^+ \bar{\nu_{\mu}} \nu_e$ were detected by 24 lead-scintillating fiber calorimeters read out by 400 MHz waveform digitizers, yielding both time and energy information.  Since this is a weak decay, the high energy positrons preferentially point in the direction of the muon spin, such that an energy threshold cut at 1.8 GeV produces a modulation in the number of positrons detected as a function of time, multiplied by the muon decay curve:
   \begin{equation}
   	N(t) = N_0 \exp( -\frac{t}{\gamma\tau} )
   		\left[ 1 + A \sin( \omega_a t + \phi_a ) \right]
   \end{equation}   
where A (or Asymmetry) is the depth of the modulation  and $\tau$ is the muon lifetime at rest.  Figure \ref{wiggle} gives a semi-log plot of the modulation curve from the 2001 g-2 data covering almost 9 muon lifetimes. 

This form was modified by beam dynamics, pileup, gain corrections at early times, and muon losses coming from processes other than decay.  Differences in the way in which each of these effects was treated, as well as data selection and pulse finding, resulted in four independent analyses of $\omega_a$ for the 2000 data \cite{g202} and five for the 2001 data \cite{g204}, which were then averaged, with attention to their correlated uncertainties.

\begin{figure}[tbp]
\centering
\includegraphics[width=3.0in]{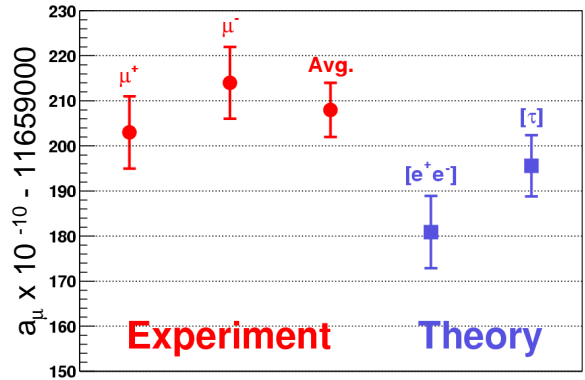}
\caption{ Comparison of a$_{\mu}$(exp) from the $\mu^-$(2000) and $\mu^+$(2001) BNL runs with a$_{\mu}$(SM) using a$_{\mu}$(Had) from both e+e- and $\tau$-decay parameterizations.}     
\label{latestexp}
\end{figure}

The analysis of $\omega_a$ and $\omega_p$ was divided into separate tasks with secret offsets for self-blinding. The value of a$_{\mu}$ was determined after the analyses of $\omega_a$ and $\omega_p$ had been finalized, the offsets removed, and radial E-field and pitch corrections applied. The result of the last two highest precision runs are shown in figure \ref{latestexp} together with the SM result for e+e-.  The SM calculation which uses $\tau$-decay to obtain a$_{\mu}$(Had) is shown in the figure as well, in order to illustrate the degree of discrepancy between these two methods.  Assuming CPT, we combined the results from $\mu^-$ and $\mu^+$ runs to obtain a$_{\mu}$(exp) = 11 659 208(6) $\times$ 10$^{-10}$ \cite{g204}.

\section{Current Theoretical Status}	
\begin{figure}[tbp]
\centering
\includegraphics[width=4.0in]{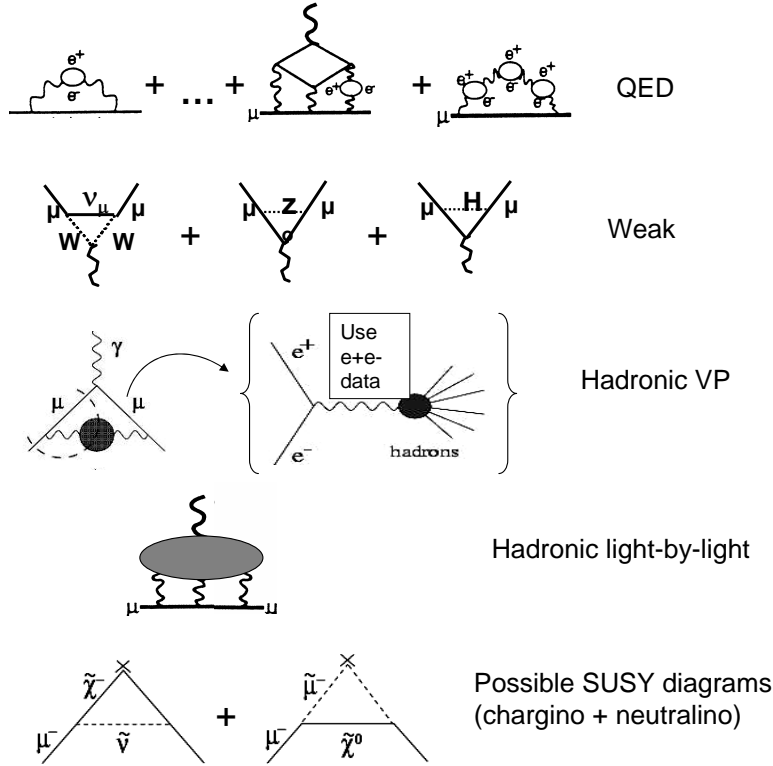}
\caption{A subset of the Feynman diagrams relevant to Standard Model calculation of the anomalous magnetic moment of the muon.  Unlike cross section experiments, measuring the g-2 frequency is sensitive to the simple sum of a$_{\mu}$ contributions, rather than the square.}  
\label{diagrams}   
\end{figure}

One can separate the components of the standard model calculation into the contributions from electromagnetic interactions, those involving weak bosons, the hadronic vacuum polarization, and the hadronic light-by-light scattering.  Such a sum is shown in figure \ref{diagrams} using a subsample of Feynman diagrams. The hadronic vacuum polarization contribution cannot be calculated from perturbative QCD,  but instead must be related to the measured hadron production cross section R(s) in e+e- collisions via the dispersion relation given by equation 1. This is graphically illustrated in the third line of figure \ref{diagrams} by a dotted line cutting the virtual hadronic blob to demonstrate how the real process of e+e- to hadrons is related to the virtual process which must be parametrized. The dominant SUSY diagrams are also included to illustrate how new physics might enter into the sum of contributions. 

The best set of such contributions, representing the latest compilations at the time of this conference, is listed below, which when added together give a$_{\mu}$(SM) $\times$10 $^{-10}$:

\vspace{.5 cm}
\noindent
a$_{\mu}$(QED) = 11658471.958  ~(0.143)              ~~~~~~~\cite{KN04}\\
a$_{\mu}$(Weak) = ~~~~~~~~~15.4    ~~~~(0.2)      ~~~~~~~~~\cite{CMV03}\\
a$_{\mu}$(Had-LO) = ~~~~693.4    ~~~~(5.3)(3.5)   ~~~\cite{HMNT04}\\
a$_{\mu}$(Had-NL) = ~~~~~~-9.8  ~~~~(0.1)        ~~~~~~~~~\cite{HMNT04}\\
a$_{\mu}$(Had l-b-l) = ~~~~13.6   ~~~~(2.5)       ~~~~~~~~~\cite{MV03}\\

The QED component dominates, but also has the smallest error (now computed up $\alpha^4$ with an estimation of $\alpha^5$).  An improved value for the $\alpha^4$ QED term and later corrections (December 2004) by Kinoshita and Nio \cite{KN04} are included.  The weak contribution includes 2-loop, leading and next-leading log, but hasn't changed much in the last decade. 

The largest error is in the first order hadronic vacuum polarization contribution as discussed above.  Although higher order calculations rely on the same parameterization, the contribution itself is much smaller and the error does not dominate. Calculations of the vacuum polarization contribution using vector spectral functions from hadronic $\tau$-decays \cite{DEHZ03} gives a contribution that differs significantly from the e+e- determination, dominated by Novosibirsk CMD-2 data. Since it also differs from recent KLOE  and BaBar results which use radiative return to reduce the center of mass energies to those most relevant to g-2, it is no longer used in direct comparisons as it requires assumptions about CVC, isospin corrections, electroweak symmetry breaking, and the charged-$\rho$ mass.  The hadronic VP contribution chosen here therefore does not include $\tau$-decay data, but does include the newest KLOE results \cite{KLOE04}, as well as a QCD fit at the higher center of mass energies.

The next largest uncertainty comes from the light-by-light term which is a model-dependent calculation.  The listed contribution and its uncertainty come from the recent re-evaluation of this term by Melnikov and Vainshtein \cite{MV03}.  A smaller overall light-by-light term (12.0(3.5) $\times$ 10$^{-10}$ due to the inclusion of some negative contributions) has been evaluated by Davier and Marciano \cite{DavMar04}, which increases the $\Delta$a$_{\mu}$ discrepancy to 2.7 $\sigma$ significance.  However, using the numbers quoted above, the g-2 discrepancy is $\Delta$a$_{\mu}$ = 23.5 (9.0) $\times$10 $^{-10}$  representing a 2.6 $\sigma$ significance. 

\section{ Muon g-2 Constraints on SUSY Dark Matter}

If supersymmetry is responsible for the non-standard part of the g-2 anomaly, there exist new diagrams which can contribute to a$_{\mu}$, specifically two new one-loop diagrams: one with an internal loop of smuons and neutralinos and one with a loop of sneutrinos and charginos (see figure \ref{diagrams}).  For minimal supersymmetry (MSSM) parameter space in the limit of large tan $\beta$, it is the chargino contribution that can most easily generate masses large enough to explain the discrepancy.  A fairly generic result for tan $\beta~>$ 5 is an inverse quadratic dependence on the SUSY loop mass given by 

\begin{equation}
|a^{SUSY}_{\mu}|=13\times10^{-10}\left[\frac{100GeV}{m_{SUSY}}\right]^2tan \beta 
\end{equation}
where tan $\beta$ is the ratio of vacuum expectation values of the Higgs doublet \cite{CzM01}.

\begin{figure}[tbp]
\centering
\includegraphics[width=4.0in]{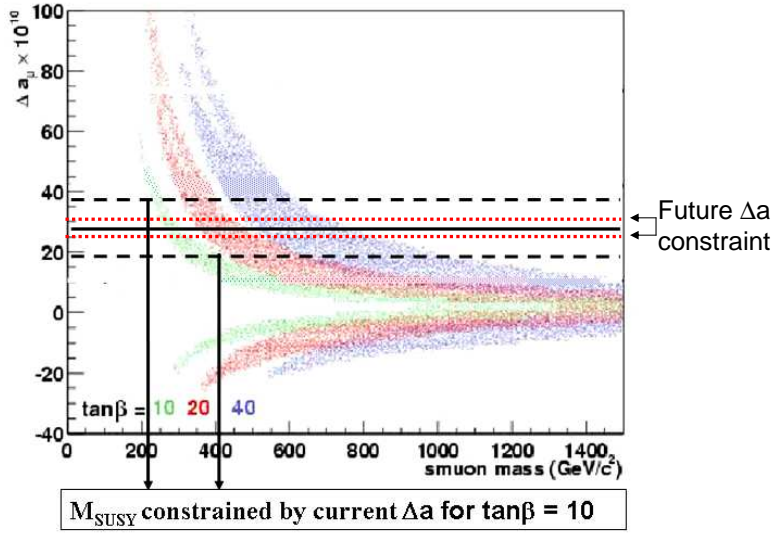}
\caption{ The shaded regions are the allowed values for the supersymmetric contribution to a$_{\mu}$ as a function of the left-handed scalar muon mass for minimal supergravity (based on plots from T.~Goto \protect\cite{Goto99}. Contraints from the Higgs boson search are already imposed. Three different tan $\beta$ values (10, 20, and 40) are shown.  The mean and 1-$\sigma$ bounds of $\Delta$a$_{\mu}$from the most recent experiment and SM calculation provide a straight line from which to determine m$_{SUSY}$ limits. }
\label{goto}     
\end{figure}

\begin{figure}[tbp]
\centering
\includegraphics[width=3.5in]{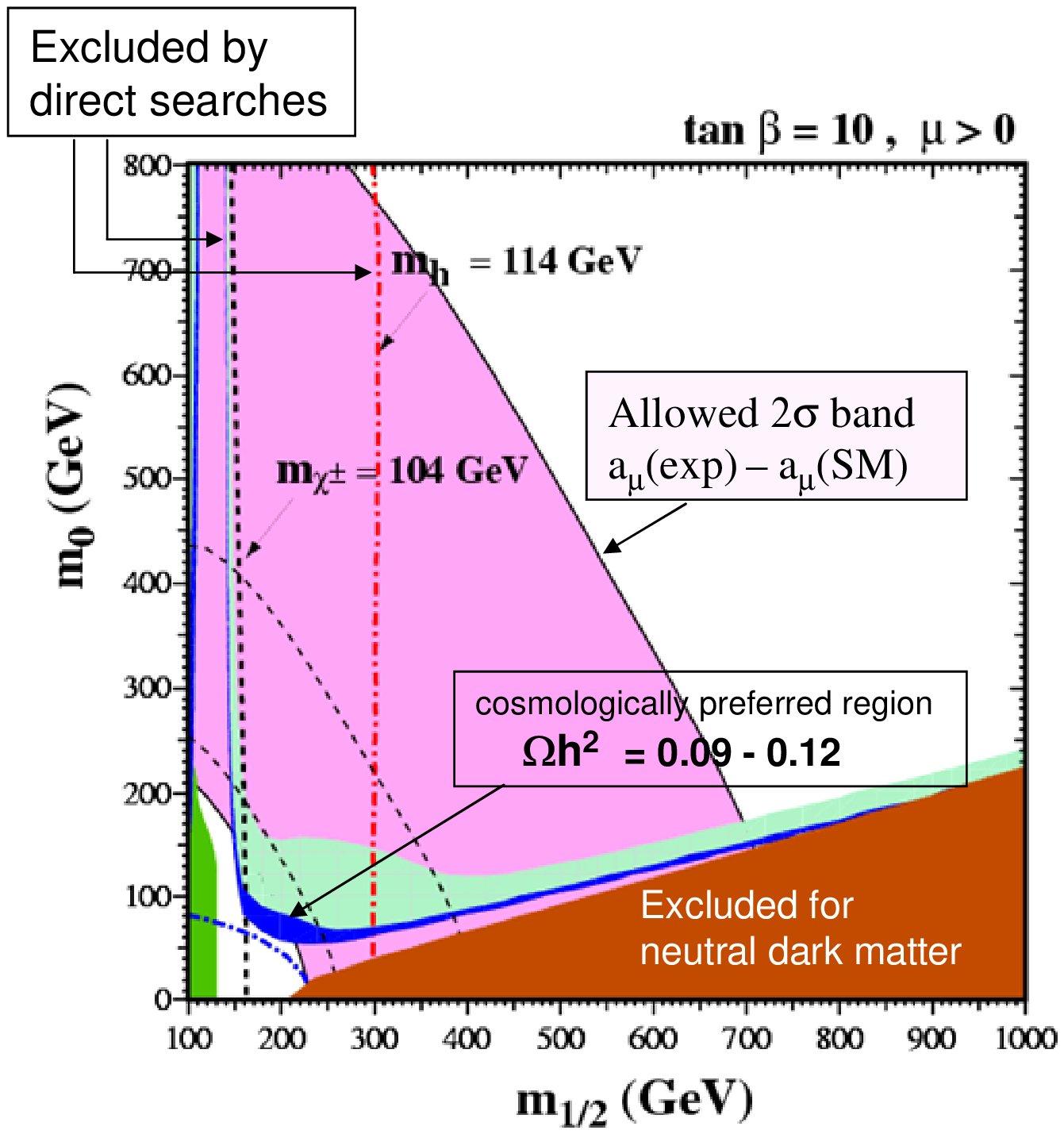}
\caption{Courtesy of Keith Olive. The m$_{1/2}$  vs m$_0$ planes in CMSSM.  The cosmologically-preferred  region allowed by  WMAP constraint (0.094 $< \Omega_{CDM}h^2<$ 0.129) is the thinner dark boomerang. The shaded region is favored  by g-2 at the 1$\sigma$ level (dotted lines) and 2$\sigma$ level (solid lines)}  
\label{Olive}   
\end{figure}

This (m$_{SUSY}$)$^{-2}$ dependence is responsible for shapes of the shaded regions in the plot of figure \ref{goto}, provided by T. Goto \cite{Goto99} for three tan $\beta$ regions, under the minimal supersymmetric extension of the Standard Model and the framework of SU(5) GUT models.  The current $\Delta$a$_{\mu}$ value (solid line) and its 1-$\sigma$ bounds (dotted lines) are plotted on top, with vertical arrows to show how m$_{SUSY}$  is limited by the g-2 experiment for a particular value of  tan $\beta$  (tan $\beta$=10).  When such constraints are translated into a 2-D plot of gaugino (m$_{1/2}$) vs scalar (m$_0$) mass \cite{EOSS03} in the constrained minimal supersymmetric model (CMSSM), they form the quarter circle shape of the g-2 preferred mass region.  Figure \ref{Olive} shows such a plot for a particular choice of tan $\beta$  and $\mu$ (the Higgs mixing parameter).  The dotted lines represent the 1-$\sigma$contours and the solid lines bounding the shaded region correspond to the 2-$\sigma$ contours on a g-2 discrepancy presumed to be saturated by the SUSY contribution. For this plot, Olive has used the value of $\Delta$a$_{\mu}$ = (23.9 $\pm$ 9.9) $\times$10$^{-10}$ corresponding to the theory compilations in March 2004, but not very different from the current value quoted here. As tan $\beta$  is increased the quarter circle stretches and moves to higher mass.  Both the positive nature of the g-2 discrepancy and the b$\rightarrow$s$\gamma$ branching ratio constraint prefer positive $\mu$.

The power of the g-2 measurement to constrain SUSY dark matter lies in the contrasting way in which it cuts across m$_0$ vs m$_{1/2}$ parameter space compared to the cosmologically preferred region, which is a hyperbolically thin dark line with co-annihilation strips  extending to high m$_{1/2}$  and m$_0$.  Connecting these two high-mass extensions is the central "focus point", considerably shrunk from the fatter (light-shaded) region by the WMAP data (0.094 $< \Omega_{CDM}h^2<$ 0.129).   LEP data excludes low m$_{1/2}$ regions.  The requirement that the dark matter particle be neutral eliminates the lower right triangle where the stau becomes the lowest mass SUSY particle. The next generation of collider searches will take place at CERN when the Large Hadron Collider comes on line in summer of 2007.  The narrower the g-2 band, the more tan $\beta$ itself will be constrained if supersymmetric particles are discovered and their mass measured at a collider.  

\section{Future Experiments and Theory Advances}
Improvement in the g-2 constraint will depend on future advances in theory and whether or not a new g-2 experiment can be mounted at Brookhaven in the near future.  Only 20\% of the CMD-2 e+e- data (center of mass energies from 0.3–1.4 GeV) have been analyzed.  Over the next several years one should expect the precision in the dispersion integral to improve as this work is completed.  An upgrade to the VEPP-2000 collider will provide increased luminosity and an improved detector (SND) to add statistics to the Novosibirsk data sample.  An intensity upgrade at the BEPS machine will increase the sample of e+e- data at the intermediate 2-5 GeV range.  This energy range contributes less to the g-2 hadronic correction, since the kernel K(s) is decreasing with s, but it does provide an important handle on potential systematic bias in the region where is overlaps CMD-2 and previous experiments. BaBar, KLOE and Belle will weigh in on differential cross sections using radiative return for multiple pion states. The BaBar data will be especially interesting, since the data can be directly normalized in the same apparatus measuring e+e- $\rightarrow \mu+\mu-$.  Within the decade, the error on the 1$^{st}$ order hadronic correction should be reduced to $\delta$a$_{\mu}\sim $35 $\times$ 10$^{-11}$, which is comparable to the uncertainty on the hadronic light-by-light contribution. Since hadronic light-by-light  scattering is model-dependent, it is hard to predict whether a breakthrough will occur there anytime soon.  Lattice gauge calculations may have some successes in the next few years. 

\begin{figure}[tbp]
\centering
\includegraphics[width=4.5in]{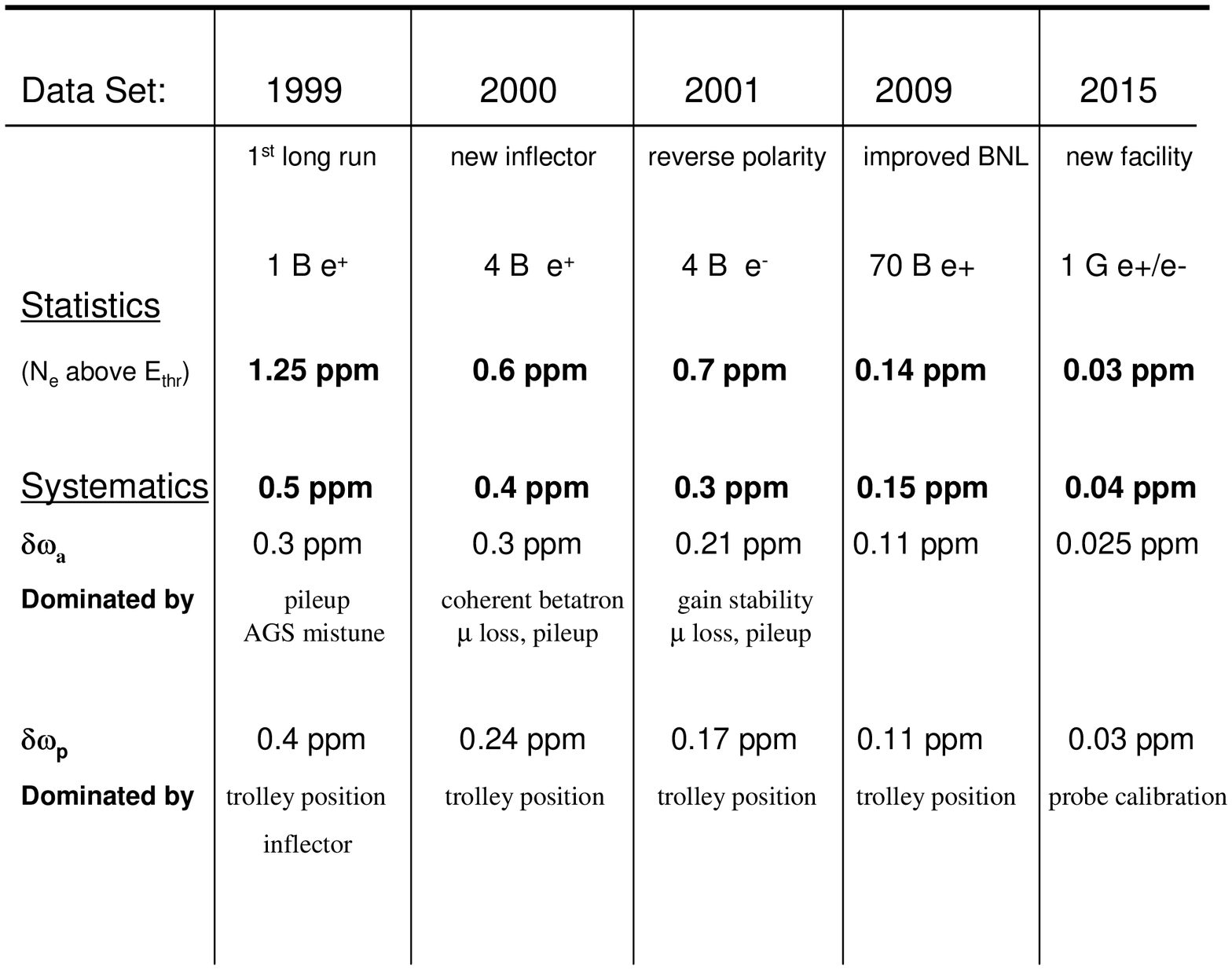}
\caption{Evolution of the statistical and systematic uncertainties in the BNL g-2 experiment. The first three columns refer to completed experimental runs.  Only the most recent g-2 runs have been included. Note that the experiments have all been statistically limited. The last two columns represent future experiments: a BNL experiment in the near future with modifications to the existing beamline, storage ring and detectors, and a possible second generation experiment to be staged at JHF in Japan. }
\label{table}     
\end{figure}

On the experimental side, Figure \ref{table} shows the evolution of the measured g-2 precision.  It can be seen that each run is statistics limited and that the systematic uncertainties for $\omega_a$ and $\omega_p$ are comparable. Another run in 2009 as E969 will represent the best one can do with the modified ring and detector geometry, before becoming limited by systematics. This requires collecting 70 billion decay positrons.  By doubling the number of beamline quadrupoles and using an open-ended inflector design, the number of stored muons can be increased by a factor of 5, allowing this to be done in  only 21 weeks for a 0.14 ppm statistical error.  The systematic error on $\omega_a$ can be reduced by injecting backward-going muons to reduce pion flash, adding another  kicker module to reduce coherent betatron oscillations,  segmenting calorimeters to reduce rate-dependent effects, and improving the front end electronics and data acquisition to handle the increased throughput.  In situ measurement of the kicker eddy currents and mapping of the NMR probes can reduce $\delta\omega_p$.  Combined with the theory precision expected a few years from now, the error on $\Delta$a$_{\mu}$ would then be at 4.7 $\times$ 10$^{-10}$.   If the mean $\Delta$a$_{\mu}$ remains stable, this represents a close to 6$\sigma$ departure from the Standard Model.

Looking beyond BNL, the next generation g-2 experiment would need a factor of 100 more data to make it worthwhile. Such concepts are being explored at the JPARC facility in Japan, (see, for example, Miller \cite{Miller02}) where JHF provides a factor of 10 increase in intensity (100 bunches/cycle every 0.7 ms) and the rest would have to come from an improved match between beam line and storage ring, etc.  Another way to improve the experiment would be to increase the energy of the muons (and their dilated lifetime) so that more g-2 cycles can be measured for the same number of stored particles.  This means abandoning electrostatic focusing, which can only be used at 3.1 GeV, that "magic" momentum where the radial electric field term cancels and the precession is unaffected. A new ring structure has been proposed by Farley \cite{Farley03} which replaces electrostatic quads with edge focusing.  Due to the large inhomogeneities in the field, the NMR probes must be replaced by proton calibration of the field.  All such new initiatives are still more than a decade away.

\begin{figure}[tbp]
\centering
\includegraphics[width=4.5in]{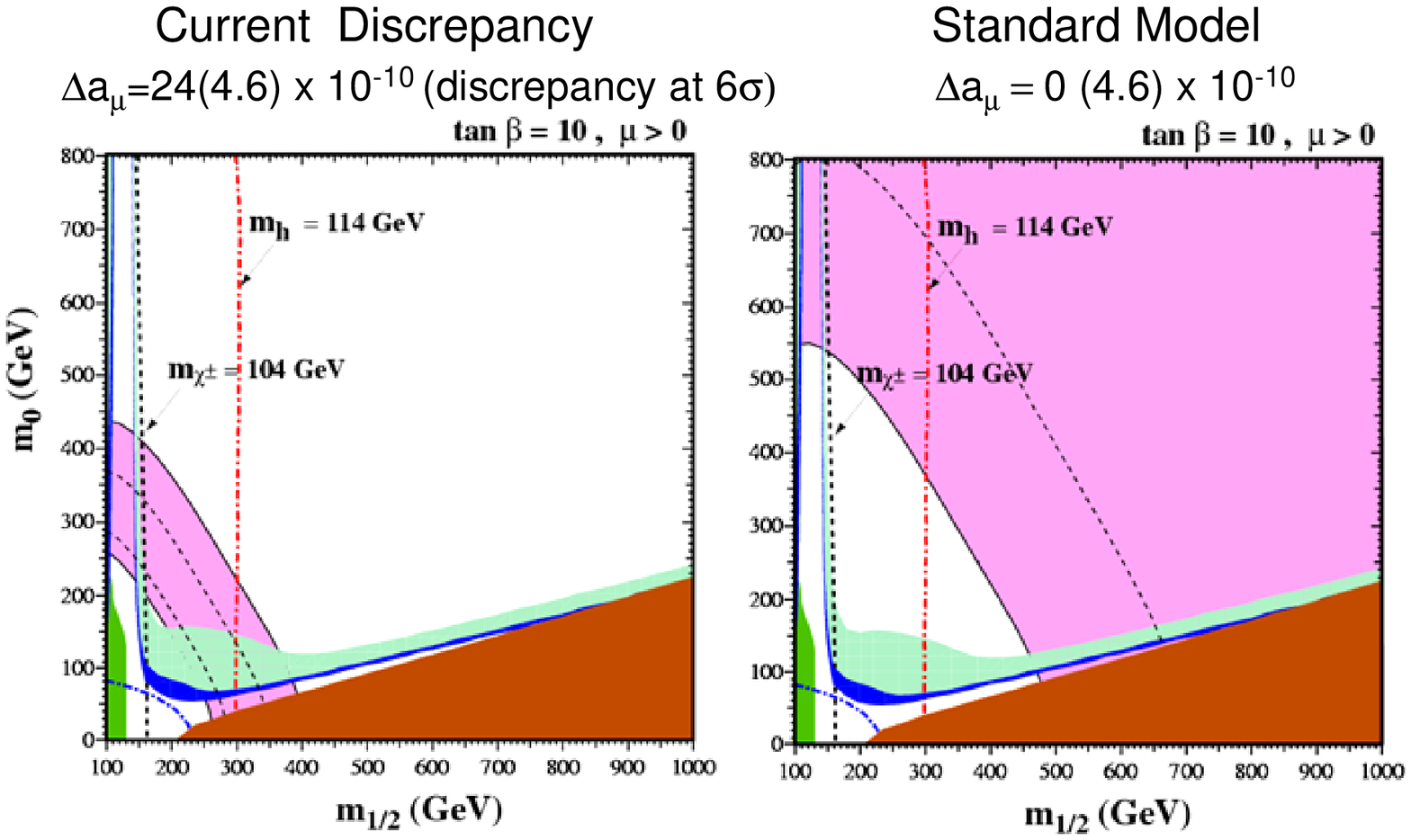}
\caption{The CMSSM m$_{1/2}$  vs m$_0$ planes for tan $\beta$=10, $\mu >$0 as in figure \ref{Olive} (Courtesy of Keith Olive). However, here the g-2 allowed regions correspond to two possible future scenarios, each with the reduction in uncertainty expected from a new E969 g-2 experiment and an improved hadronic VP correction as e+e- statistics increase. Left plot: Mean discrepancy remains the same to give $\Delta$a$_{\mu}$ = 239 (47) $\times$ 10$^{-11}$.  Right plot: The experiment matches the standard model value to give $\Delta$a$_{\mu}$ = 0 (47) $\times$ 10$^{-11}$. }
\label{Olive2}     
\end{figure}

Figure \ref{Olive2} shows how reduced errors would translate into dark matter constraints for a particular choice of tan $\beta$=10 and the preferred $\mu>0$. Both plots include the factor of 2.5 reduction in uncertainty expected from a new run at BNL, combined with improvements in a$_{\mu}$(Had) from e+e- data already collected.  The plot at the left represents the case where the mean discrepancy remains stable at its present value.  The plot to the right represents the case where the mean shifts down to the SM value.  Due to the nature of the constraints, a reduction in the error bars which leaves the mean $\Delta$a$_{\mu}$ intact will significantly narrow the band of allowed masses, while a shift down to SM will widen the allowed region, but reject SUSY masses $<$ 500 GeV/c$^2$.  In both cases, this will provide significant new constraints on SUSY dark matter under the CMSSM.

\section{Conclusions}
On the experimental front, the BNL g-2 experiment has succeeded in its goal to improve the precision of a fundamental constant by a factor of 15 since the last CERN experiment 30 years ago.  However, the theoretical landscape has shifted considerably.  Originally, the BNL experiment was designed to search for the Higgs and to confirm electroweak symmetry-breaking by measuring a$_{\mu}$(weak) to 20\%.  As the mass limits on the Higgs moved upward over the last decade and a half of Tevatron and LEP runs, the contribution to a$_{\mu}$ from diagrams containing the Higgs shrank below our sensitivity.  The popularity of SUSY as an answer to the hierarchy problem and as a means to unify gauge couplings has renewed interest in a$_{\mu}$, especially since the hint of a discrepancy points to such convenient SUSY masses.  The experimental improvement has lead theorists to uncover a number of errors, improve calculations involving both hadronic vacuum polarizations, as well as higher order QED terms, spurred further experimental work on R(s), and lead to a re-examination of CVC and pion form factors.  In the end, no matter what the fad of the moment, precision measurements of fundamental constants are an enduring contribution to physics, since they confront our preconceptions with reality and guide future discussions.

\printindex
\end{document}